\documentclass[aps,prl,twocolumn,groupedaddress,floatfix,showpacs,nofootinbib]{revtex4-1}
\usepackage{amssymb,amsmath}
\usepackage{graphicx}
\usepackage{subfigure}
\usepackage[english]{babel}
\usepackage{float}
\usepackage{color}
\usepackage[symbol]{footmisc}

\newcommand{\fig}[1]{Fig.~\ref{#1}}
\newcommand{\figs}[1]{Figs.~\ref{#1}}

\begin{document}

\title{Controlled transport based on multiorbital Aharonov-Bohm photonic caging}

\author{Gabriel C\'aceres-Aravena$^{1,2,\dagger}$, Diego Guzm\'an-Silva$^{1,2,\dagger}$, Ignacio Salinas$^{1,2}$, and Rodrigo A. Vicencio$^{1,2}$}
 
\affiliation{$^{1}$Departamento de F\'isica, Facultad de Ciencias F\'isicas y Matem\'aticas, Universidad de Chile, Chile
\\ $^{2}$Millenium Institute for Research in Optics - MIRO, Universidad de Chile, Chile}

\date{\today}



\begin{abstract}
The induction of synthetic magnetic fields on lattice structures allows to effectively control their localization and transport properties. In this work, we generate effective $\pi$ magnetic fluxes on a multi-orbital diamond lattice, where first ($S$) and second ($P$) order modes effectively interact. We implement a $z$-scan method on femtosecond laser written photonic lattices and experimentally observe Aharonov-Bohm caging for $S$ and $P$ modes, as a consequence of a band transformation and the emergence of a spectrum composed of three degenerated flat bands. As an application, we demonstrate a perfect control of the dynamics, where we translate an input excitation across the lattice in a completely linear and controlled way. Our model, based on a flat band spectrum, allows us to choose the direction of transport depending on the excitation site or input phase.
\end{abstract}

\maketitle

Understanding the properties of transport and localization in lattice systems has been crucial for the implementation of present technologies based in electronics. Novel phenomena has been discovered where, for example, the transport can be drastically reduced and localization effects could emerge. One important example is the well known Anderson localization~\cite{Anderson}, where all the eigenmodes collapse due to a destructive interference process associated to a random distribution of the site energies and/or lattice positions~\cite{ALmoti,ALmora}. Another interesting and more controllable method appears with the addition of a magnetic flux on a lattice, which adds an extra phase on hopping (coupling) coefficients. For very specific values of this flux, a phenomenon of extreme localization is observed, the so-called \textit{Aharonov-Bohm (AB) caging}~\cite{ABcage,Vidal98}. In this case, in contrast to the Anderson localization, all the eigenmodes become degenerated and form a set of flat bands (FBs)~\cite{FBRV}. Therefore, an input excitation is decomposed in FB compact states, the energy is caged, and the transport is abruptly reduced into a couple of unit cells. Although the use of a magnetic flux was initially thought for electronic lattices, it was possible to map its effect to non-electronic systems by the use of artificial gauge fields~\cite{gaugefields}. This opened the possibility to study the AB effect in other physical systems~\cite{Longhi14,Fang12,Bloch12,Brosco21,Perrin20,miller12,Bermudez11}, as well as to experimentally observe it in superconducting wire networks~\cite{Abilio99} and in semiconductors tailored in an electron gas~\cite{Naud01}. Very recently, a synthetic magnetic flux has been implemented in photonics by applying a detuning and modulation of curved waveguides~\cite{Sebabrata}, by the addition of auxiliary waveguides to generate an effective negative coupling~\cite{Kremer20}, or by the excitation of special beams having orbital angular momentum~\cite{Jorg21}.

While the AB effect allows to trap and steer the propagation of light on a given region, other ways of controlling the dynamics in optical lattices have been suggested as well~\cite{GARANOVICH12}. When studying mobility properties to, for example, modify the direction of propagation of a light beam, most of the efforts have been focused on the addition of nonlinearity to the system~\cite{GARANOVICH12,Molina06,Naether11,cristian13,Peschel02}, which also induces instabilities and resonances with linear modes. Interestingly, magnetic fluxes in non-Hermitian lattices have also the property to alter the direction of propagation affecting both reflexion and transmission, and producing unidirectional reflection, absorption and lasing~\cite{Jin16,Regensburger12,Feng13,Ramezani14,Ramezani16,Jin18}. 
However, the control of localized wavepackets, including a magnetic flux in Hermitian systems, has not been experimentally observed so far. A controlled flow of energy is important in general physics and also when thinking on concrete applications as, for example, logical operations and concatenated optical gates in photonics and integrated technologies~\cite{IPperez,IPose}. 
%
\begin{figure}[t!]
\centering
\includegraphics[width=0.92\columnwidth]{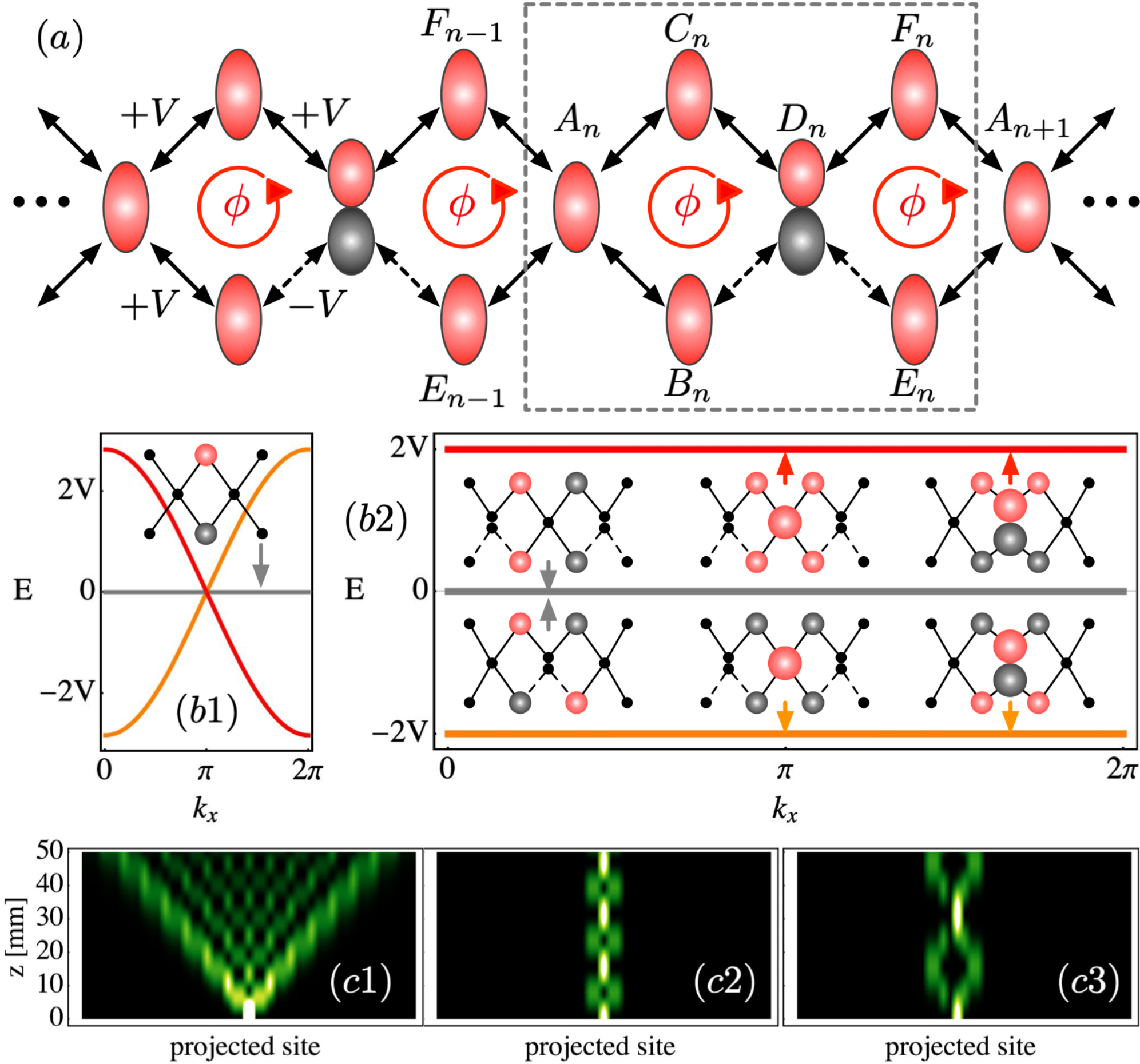}
\caption{(a) A multi-orbital diamond lattice formed by $S$ and $P$ sites. Full (dashed) arrows indicate a positive (negative) coupling constant. The effective induced magnetic flux per plaquette is denoted as $\phi$. A dashed rectangle defines the unit cell formed by six lattice sites. (b1) and (b2) Spectrum for $\phi=0$ and $\phi=\pi$, respectively. FB profiles are shown as insets. (c) Intensity evolution for different input conditions: (c1) central site for $\phi=0$; (c2) central site and (c3) top/bottom site, both for $\phi=\pi$. In (c) $V=0.1$ mm$^{-1}$ and $z_{max}=50$ mm.}
\label{fig1}
\end{figure}

In this Letter, we propose a novel form to implement a synthetic magnetic flux in lattices via the use of inter-orbital coupling~\cite{Diego21} on a diamond lattice~\cite{Longhi14}. By considering the effective interaction between fundamental ($S$) and second-order ($P$) modes, a positive and negative coupling can be obtained depending on waveguide and mode orientation. This allows to induce an effective magnetic flux $\phi=\pi$ per plaquette (a closed ring in the lattice), allowing a direct observation of AB caging. By using a femtosecond (fs) laser writing technique~\cite{Szameit05}, we fabricate a multi-orbital photonic lattice having a mixture of straight waveguides which support one or two modes. We observe quite clearly AB caging for $S$ and $P$ states. As a remarkable application, we experimentally demonstrate a novel mechanism for directional transport of linear localized excitations. We show a perfect steering of the energy across the lattice in a very compact way and suggest a concatenated photonic scheme. Our results are certainly a step forward on the study of discrete dynamics~\cite{rep1,rep2}, offering a new and linear solution for the mobility of localized wavepackets in lattices.

Our model consists of a multi-orbital diamond-like chain, which includes sites that support fundamental modes only and sites that support also a second mode, as it is sketched in \fig{fig1}(a). As the mathematical model is equivalent to the propagation of quantum particles on a tight-binding lattice, we use the hydrogen-like nomenclature of S and P orbitals. We have very recently demonstrated~\cite{Diego21} that an effective interaction between different orbitals can be achieved when a tuning process is implemented, such that the optical propagation constants (or the energies in a quantum version) match. This also means that the $S$ mode of tuned $P$ sites is not anymore in resonance with the $S$ mode of $S$ sites and, therefore, they simply do not interact. The unit cell of this lattice consists of six sites as shown by a dashed rectangle in \fig{fig1}(a). The coupling interaction between sites is determined by the coupling constants $\pm V$, which depends on the spatial orientation of waveguides and their respective modes. We have defined the positive part of the wave functions using red colors, while the negative part is in gray. For any distance, the horizontal coupling in between $S$ and $P$ waveguides, at the central row, is zero due to the orthogonality of both wavefunctions and the $\pi$ phase structure of the $P$ mode~\cite{Sutton93}. In fact, this makes possible to achieve positive and negative couplings constants on a given lattice geometry and induce new phenomena on rather trivial systems, without the need of implementing more complex configurations~\cite{Sebabrata,Kremer20,Jorg21}. 

A single negative coupling constant on a plaquette effectively induces a synthetic flux of $\phi=\pm\pi$~\cite{Longhi14}, which is the main requirement for the observation of AB caging. We use a tight-binding approach to write down the coupled equations governing the dynamics of this system~\cite{SM}, which in a compact form can be written as
\begin{equation}
-i\frac{\partial \psi_n}{\partial z}=\sum_{m} V_{n,m}\psi_m\ .
\label{eq}
\end{equation}
Here, $\psi_{n}(z)$ describes all the lattice amplitudes, $z$ the propagation distance (dynamical coordinate), and $V_{n,m}$ the nearest-neighbor coupling elements. First of all, we look for the lattice spectrum by assuming a Bloch ansatz of the form $\psi_{n}(z)=\psi_{0} \exp\{i k_x n\}$, with $\psi_{0}$ a site amplitude and $k_x$ the horizontal quasi-momentum. We notice that for a homogeneous lattice, composed of $S$ waveguides only, the effective flux is simply $\phi=0$, the unit cell reduces to only three sites, and the spectrum has just three bands~\cite{Longhi14}: $E=0,\ \pm2\sqrt{2}\cos(k_x/2)$. We plot this spectrum in \fig{fig1}(b1) and observe a FB with compact localized states~\cite{FBRV,SebaDia} as the ones shown at the inset. FB localization is typically achieved by destructive interference of positive and negative amplitudes at specific connector sites~\cite{FBRV,fbluis} which, in this case, correspond to the ones at the central row. Therefore, the excitation of central sites does not resonate with FB states and excites dispersive bands only (more details in~\cite{SM}), as the numerical propagation shows in~\fig{fig1}(c1).

On the other hand, a multi-orbital lattice as the one shown in~\fig{fig1}(a) induces an effective flux of $\pi$ at closed plaquettes. This small modification on coupling coefficients produces a fundamental change on the linear properties: all the linear modes become compact and completely localized. The synthetic magnetic field cancels the lattice dispersion and no extended eigenmodes are allowed to exist in the system. The linear spectrum has six bands with only three different eigenvalues (each band having a two-fold degeneracy): $$E_0=0\ ,\ \ E_+=2V\ ,\ \ E_-=-2V\ .$$ This flat spectrum is shown in \fig{fig1}(b2), including the six different FB states as insets. We immediately notice that a single-site excitation will excite a reduced number of FB states only. The AB caging is usually understood~\cite{Sebabrata,Kremer20,Jorg21} as a symmetric oscillation of the energy on a narrow lattice region, as a consequence of exciting a couple of localized FB states corresponding to two different energies. When injecting light at sites $S$ or $P$ at the central row, only two FB states will be excited with energies $\pm2V$, with a caging period of $z_c=\pi/2V$~\cite{SM}. We simulate this scenario in~\fig{fig1}(c2) where a central row site ($S$ or $P$) is excited. We observe how the energy is trapped and goes back and forward, oscillating around the input position; i.e., \textit{the energy is caged}. A perfect destructive interference occurs at the next central sites and no amplitude is transmitted to the rest of the lattice. As a consequence the energy returns back to the input position and a caging cycle starts again. \fig{fig1}(c3) shows a periodic dynamics also, when light is injected at top/bottom $S$ sites, as a result of exciting three FBs~\cite{SM}.

\begin{figure*}[t!]
\centering
\includegraphics[width=0.95\textwidth]{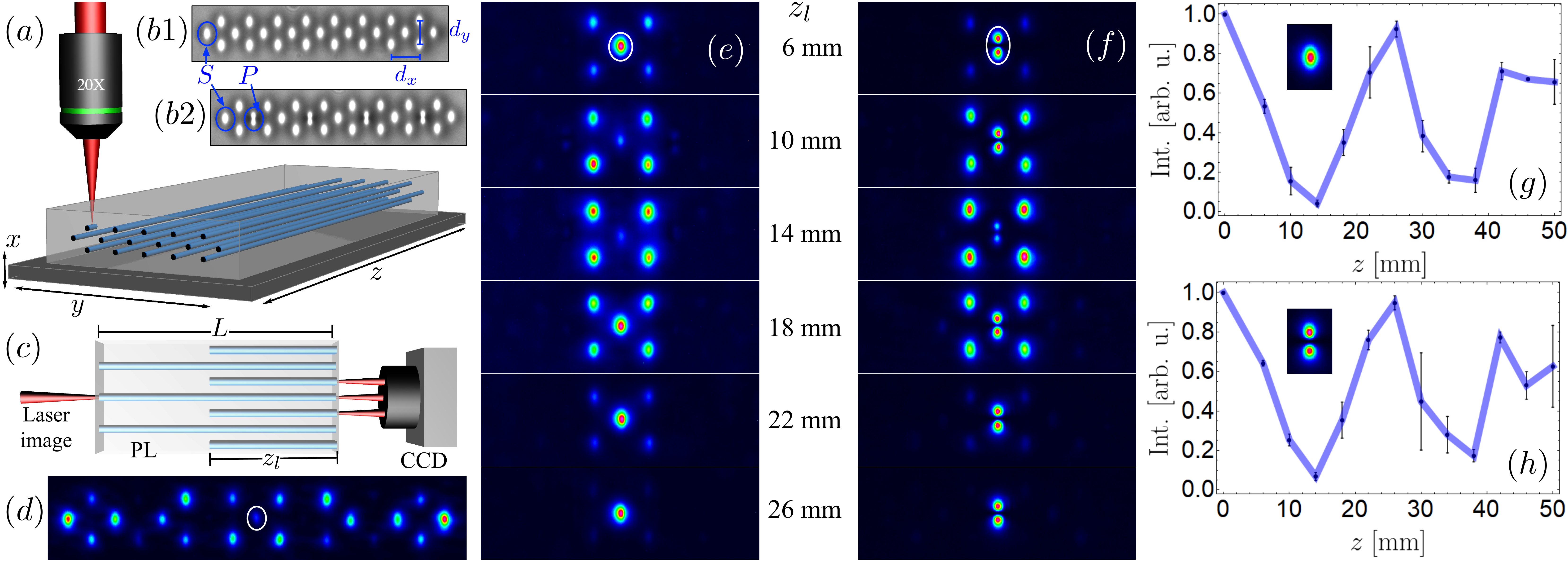}
\caption{(a) fs laser writting technique. (b1) and (b2) Bright-field microscopy images for a homogeneous and a multi-orbital diamond lattice, respectively. (c) Top view of the image characterization setup and $z$-scan concept. (d) Intensity output image at $z=L$ for $\phi=0$. (e) and (f) Intensity output images at $z=z_l$ and $\phi=\pi$, for a $S$ and a $P$ mode excitation, respectively. A circle shows the input site in (d)--(f). (g) and (h) Average intensity at the input waveguide versus propagation length $z_l$, for a $S$ and a $P$ mode excitation, respectively. Error bars correspond to the standard deviation.}
\label{fig2}
\end{figure*}

We fabricate a set of photonic lattices (PL) by using a fs laser writing technique~\cite{Szameit05}, which is sketched in~\fig{fig2}(a). In this technique, pulses of $\sim 240$ fs and $\sim82$ mW are tightly focused inside a borosilicate glass wafer of length $L=50$ mm, while a high precision XYZ Aerotech stage translates the wafer at a velocity of $0.4$ mm/s. We start by fabricating a homogeneous diamond $S$ lattice as shown in~\fig{fig2}(b1), with nominal distances $\{d_x,d_y\} = \{36.00,29.72\}\ \mu$m. We experimentally test the lattices by using an image setup~\cite{FBlieb,SM}, which modulates in amplitude and phase a wide laser beam of $640$ nm. We image a specific input condition onto the PL input facet, as sketched in~\fig{fig2}(c), and measure the output profile using a CCD camera. For the homogeneous lattice, we excite a single $S$ waveguide and observe a well diffracted spatial profile, as shown in~\fig{fig2}(d). We notice that the energy covers all the lattice, as a result of exciting only the dispersive part of the spectrum~\cite{SM}. We increase the writing power to $104$ mW for the $P$ waveguides (four in our experiment), which are located at the central row as shown in~\fig{fig2}(b2). The writing powers were chosen to reach a maximum coupling in between $S$ and $P$ modes~\cite{Diego21}, which occurs when their propagation constants are precisely tuned~\cite{SM}. We implement a $z$-scan method to characterize the dynamics along the propagation coordinate, and fabricate twelve multi-orbital lattices. All $S$ and $P$ waveguides at the central row have a full length $L$, while the rest of waveguides at the top and bottom rows have a shorter propagation length $z_l$, from $6$ to $50$ mm in steps of $4$ mm, as sketched in~\fig{fig2}(c). In this way, by exciting a single $S$ or $P$ waveguide at the central row, we are able to trace the dynamics at different distances. 

First of all, we excite a $S$ central site and observe quite clearly $S$ mode caging in~\fig{fig2}(e), for the distances indicated in the central column. We observe that the light is able to couple to the top and bottom nearest-neighbour sites, but it can not propagate further through the lattice due to destructive interference at the next central sites. We measure the output intensity at the input site for different $z_l$ and plot the data in~\fig{fig2}(g), where we average the data for the three central $S$ waveguides of our lattice. The light oscillates twice for $50$ mm, with a caging period of $z_c\approx 26$ mm, what implies a coupling of $V\approx 0.06$ mm$^{-1}$. The numerical simulation shown in~\fig{fig1}(c2) described an oscillation for $S$ and $P$ modes. Therefore, we should observe a similar caging phenomena also by using a $P$ mode as an input condition, which consists of two lobes with a $\pi$-phase difference. The output profiles obtained after exciting the lattice with a second mode at $P$ central waveguides are shown in~\fig{fig2}(f). We nicely and quite clearly observe AB caging for a $P$ mode. The light tries to escape from the input site and couple to its neighbour $S$ waveguides, but already having a phase difference due to the $P$ parity. This makes impossible for the light to continue escaping because of the destructive interference occurring at the next central $S$ sites. Therefore, an effective wall is created due to the induced synthetic magnetic field, and the light is reflected back into the input position, and the cycle starts again. In~\fig{fig2}(h) we observe a two cycle oscillation in $50$ mm and similar caging period for a second mode excitation (we averaged the data considering the four central $P$ waveguides). In both cases, we observe that more than $90\%$ of the energy is caged in a first cycle, with an intensity reduction after two beating lengths. As a perfect tuning and perfect coupling symmetry conditions are really hard to be achieved experimentally, we expect to have some dispersion while increasing the propagation distance. Once the destructive interference is not perfectly balanced due to any symmetry breaking mechanism, some energy will be radiated through the lattice. Nevertheless, we observe in Figs.~\ref{fig2}(e)--(h) that most of the energy is very well confined in a narrow spatial region, and that a multi-orbital caging is clearly demonstrated~\cite{Sebabrata,Kremer20,Jorg21}.

\begin{figure}
\centering
\includegraphics[width=0.9\columnwidth]{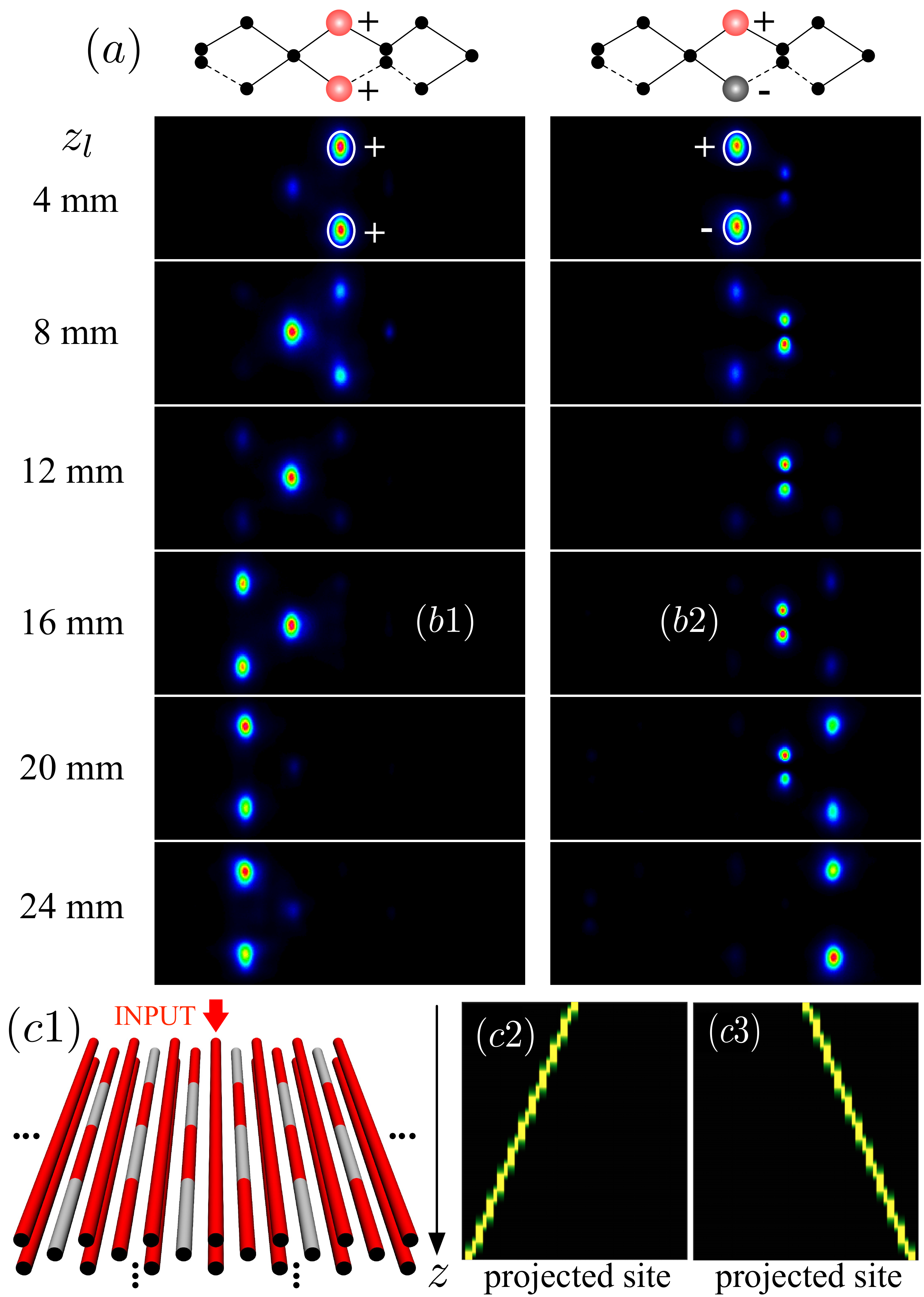}
\caption{(a) Optical mask with an in-phase (left) and an out-of-phase (right) input excitation. (b1) and (b2) Intensity output images at $z=z_l$ and $\phi=\pi$, for an in-phase and an out-of-phase input excitation, respectively. Circles show the input sites. (c1) Concatenation scheme: red and gray colors correspond to $S$ and $P$ waveguides, respectively. (c2) and (c3) Numerical propagation for in-phase and out-of-phase input conditions, respectively.}
\label{fig3}
\end{figure}
%
We notice that in a homogenous array the transport is uniformly distributed in both directions [see~\fig{fig2}(d)], while using a multi-orbital mask an unidirectional displacement could be observed, allowing us to decide if the light propagates to the right or to the left of the system. We fabricate a smaller multi-orbital lattice, which we call an \textit{optical mask}, as the one sketched in~\fig{fig3}(a). Again, we implement a $z$-scan study by fabricating twelve lattices, where the $S$ top and bottom sites of the central column have a full length $L$, while the rest lattice sites have a shorter propagation distance $z_l$, which goes from $4$ to $26$ mm in steps of $2$ mm. We expect to observe a maximum translation of the energy at $z_l\sim 26$ mm (in our mask, the effective transversal translation is similar to the one for the AB caging, as we need to move from two $S$ sites into similar $S$ sites to the right or to the left). The experiment consists on exciting simultaneously two full length $S$ sites at the center of the lattice using an image setup~\cite{SM}. An in-phase (++) excitation [\fig{fig3}(a)-left] produces a perfect cancellation of amplitudes at the right $P$ central waveguide and, therefore, zero energy flowing to the right of the mask. In contrast, the light will propagate to the left, as the first coupling occurs with the central $S$ site only, which then couples with the following (left) top and bottom $S$ sites. There, a perfect full translation is achieved with a precise control and discrete energy transportation from one unit cell into another. The energy in this case can not continue moving to the next $P$ waveguide: the energy find an interference wall and the light is reflected back again to the input position, in a dynamical picture similar to the caging one described previously. The experimental intensity images obtained in this experiment are shown in~\fig{fig3}(b1), where we observe how the light is not allowed to propagate to the right and only is able to move to the left. In fact, our experiment nicely shows how the input condition has been precisely translated to the left in a distance $z_l\approx 24$ mm. Of course, this input condition also excites a caging-like transport due to the presence of only FBs in the spectrum, and the energy oscillates in a very well defined region. For a longer propagation the energy will go back and will replicate the in-phase input profile. The out-of-phase (+-) input condition shown in~\fig{fig3}(a)-right will show an opposite dynamics. The destructive interference will cancel the amplitude at the left $S$ central site and no energy will flow to the left. However, the $\pi$-phase difference will cancel the effect of the negative coupling and the energy will flow through the $P$ central site, but keeping the initial phase difference. Therefore, when arriving to a $S$ central site to the right, the amplitude will be just zero there. The experimental images showing the out-of-phase case are presented in~\fig{fig3}(b2). 

As a concrete and very important application, we numerically show that by concatenating several optical masks, as it is schematically described in~\fig{fig3}(c1) [every mask shifts the $S$ and $P$ central waveguides periodically in the propagation direction], we achieve a controlled displacement on a larger lattice structure. A perfect translation of the energy (to the right or to the left) is obtained, as the numerical simulations in~\figs{fig3}(c2) and (c3) indicate. In fact, by changing the input phase or input position, one can reverse the direction of movement. For example, an in-phase profile could move to the left in position $\{B_n,C_n\}$ and to the right in positions $\{E_{n\pm1},F_{n\pm1}\}$. This is a very remarkable result for lattice systems~\cite{rep1,rep2}, considering that our proposal is based on linear properties only. We therefore suggest a simple and clear method for perfect energy transport, with high controllability and stability. The injection of in-phase or out-of-phase input conditions allows to propagate the energy freely through the lattice; while, on the other hand, the injection of $S$ or $P$ states at the central row induces AB caging. Therefore, our lattice model gives a perfect linear solution for controlled transport and localization.  

In conclusion, we have proposed a new scheme for inducing synthetic magnetic fields in lattice systems. This effect relies in the interaction between fundamental $S$ and excited $P$ modes, observing a net flux of $\pi$ per plaquette. We experimentally proved this system by fabricating multi-orbital diamond photonic lattices and we observed Aharonov-Bohm caging for both modes, which is a direct experimental proof for the induction of an effective magnetic field. We demonstrated the possibility of controlling the direction of transportation determined by the input condition, which could be useful, for example, in the creation and manipulation of logic gates~\cite{IPperez,IPose}. Our proposal gives a simple on demand solution for localization and transport in lattices. We believe that our multi-orbital configuration has the flexibility to include more effects in a following study, such as non-hermiticity~\cite{zhang20} or nonlinear effects~\cite{Chang21}, in a very closed and direct form.

The authors acknowledge stimulating discussions with Magnus Johansson and Bastian Real. This work was supported in part by Millennium Science Initiative Program ICN17$_-$012, and FONDECYT Grants 1191205 and 3190601.

\dag\ Both authors contributed equally to this work

\bibliography{biblio}

\end{document}